# Generating a Family of Byzantine-Fault-Tolerant Protocol Implementations Using a Meta-Model Architecture

Graham NC Kirby, Alan Dearle & Stuart J Norcross
*School of Computer Science, University of St Andrews, St Andrews, Fife KY16 9SX, Scotland*
*{graham, al, stuart}@cs.st-andrews.ac.uk*

## Abstract

*We describe an approach to modelling a Byzantine-fault-tolerant distributed algorithm as a family of related finite state machines, generated from a single meta-model. Various artefacts are generated from each state machine, including diagrams and source-level protocol implementations. The approach allows a state machine formulation to be applied to problems for which it would not otherwise be suitable, increasing confidence in correctness.*

## 1. Introduction

The finite state machine is a widely used abstraction for describing and reasoning about distributed algorithms [1]. Here we address the problem of developing a finite state machine formulation for an algorithm whose generality precludes its expression as a single state machine. Instead, the algorithm may be characterised as a family of related state machines, each corresponding to particular values of some parameters to the general algorithm. Although family members differ in their individual states and transitions, they share a common structure dictated by the general algorithm.

Our approach is to develop a meta-model that captures the common architecture of the family of state machines. This can be executed with chosen parameter values to generate any particular member of the state machine family. The output of the meta-model is a state machine representation, from which various concrete artefacts may be generated. These include textual state machine descriptions, state machine diagrams and specialised source-level algorithm implementations.

We describe this approach via the example of a Byzantine-fault-tolerant (BFT) commit algorithm—originally motivating the work. We think that the technique could also be applied to development of other fault-tolerant protocols, making it directly relevant to the area of architecting critical infrastructures.

## 2. Background

The motivation for this work arose during development of a particular algorithm within a distributed storage system [2]. The aim of the ASA project is to develop a resilient, logically ubiquitous storage infrastructure with the following attributes:

- data can be accessed efficiently and securely from any physical location
- data is stored resiliently
- an historical record of data is available

The requirements include the following:

- it must provide a logical file system that appears the same regardless of the physical machine from which it is accessed
- files stored in the file system must be resilient to the failure and/or malicious behaviour of individual machines
- it must provide a historical record

The ASA infrastructure provides a single distributed abstract file system, which is built on a generic distributed storage layer. For scalability, this storage layer is itself implemented on a peer-to-peer (P2P) key-based routing infrastructure. The storage layer provides resilience by replicating data and meta-data on multiple P2P nodes, and actively maintaining those replicas as nodes fail, misbehave or leave the P2P overlay.

The aspect of interest here is the commit algorithm used to record a new version of a logical data item in the distributed storage layer. The algorithm is executed by all members of the set of P2P nodes on which that data item's version history is replicated. The membership of this set can change dynamically as the topology of the P2P network changes.

The purpose of the algorithm is to enable the node set to agree a global ordering of the (potentially concurrent) updates to the version history of a particular data item. The algorithm ensures that the same version history is stored on each of the replica sites. Hence, a subsequent query over the history will yield a consistent response, regardless of which replica site is used.

The algorithm is also required to be BFT, meaning that it operates correctly in the face of faulty behaviour exhibited by some subset of the nodes storing the history replicas. Faulty behaviour may include responding slowly, failing completely, or arbitrary malicious actions. As is well known, the theoretical limit on all BFT schemes is that at least *3f+1* participants are needed to give tolerance to *f* failures [3]. Hence for a replication factor *r*, yielding *r* replicas of each version history, the algorithm tolerates at most *floor((r-1)/3)* faulty participants.

Background processes run to regenerate missing replicas and to replace faulty nodes, thus here the limit applies to the duration of a particular execution of the algorithm, rather than to the lifetime of the system[1]. Additional replicas need to be generated whenever the set of nodes storing replicas of a given data item is temporarily reduced. This may occur due to fail-stop faults, which are straightforwardly detected through timeouts, or due to the detection of malicious nodes. Such nodes are eventually detected with high probability using periodic cross-checks between replica nodes.

## 3. General approach

Initially, we designed a single generic algorithm that appeared to meet the requirements outlined in the previous section, parameterised by the replication factor. In an effort to gain greater insight into its operation, we then developed a finite state machine model for a selected replication factor—four, being the simplest scheme to yield a BFT algorithm. Although neither the algorithm (about 500 lines of pseudo-code) nor the state machine (33 states with 3-4 transitions from each) were especially complex, they were non-trivial. We then faced the problem that there was no strong correlation between the code and the state machine. Thus even though we were satisfied (informally) that the state machine was correct, its creation achieved little in terms of building confidence in the algorithm.

The main reason for the disparity between the state machine and the algorithm was that the former was specific to a fixed replication factor, while the algorithm was generic. The individual states in the state machine correspond to the counts of messages that have been sent and received at particular points during the algorithm's execution. The maximum values of these counts vary with the replication factor, thus the number of states in the machine also varies. By the same argument, it is not possible to construct a single state machine that captures the generic algorithm.

Our goal at this point was to unify the state machine model and the generic algorithm, by generalising the state machine in some way. The key insight that made this possible was to identify how both the state space and the state transitions were determined by the replication factor. The state space was defined straightforwardly by the various combinations of the possible message counts, themselves bounded by the replication factor. For transitions, the important point was that some denoted simple increments in message counts, whereas others denoted actions to be performed (termed *phase transitions*). By identifying where in the state diagram phase transitions occurred, and relating these to the replication factor, it was possible to produce a generic description defining a family of related state machines. We then proceeded as follows:

- We developed a meta-model that captured the common structure among the members of the state machine family.
- We executed the meta-model with a replication factor of four to generate an abstract representation of a specific state machine, which we then checked for consistency with the original state machine.
- Once satisfied with the correctness of the meta-model, we developed tools to generate various state machine artefacts, including diagrams and source-level implementations.

### 3.1. Generation process

The overall generation process is illustrated in Figure 1.

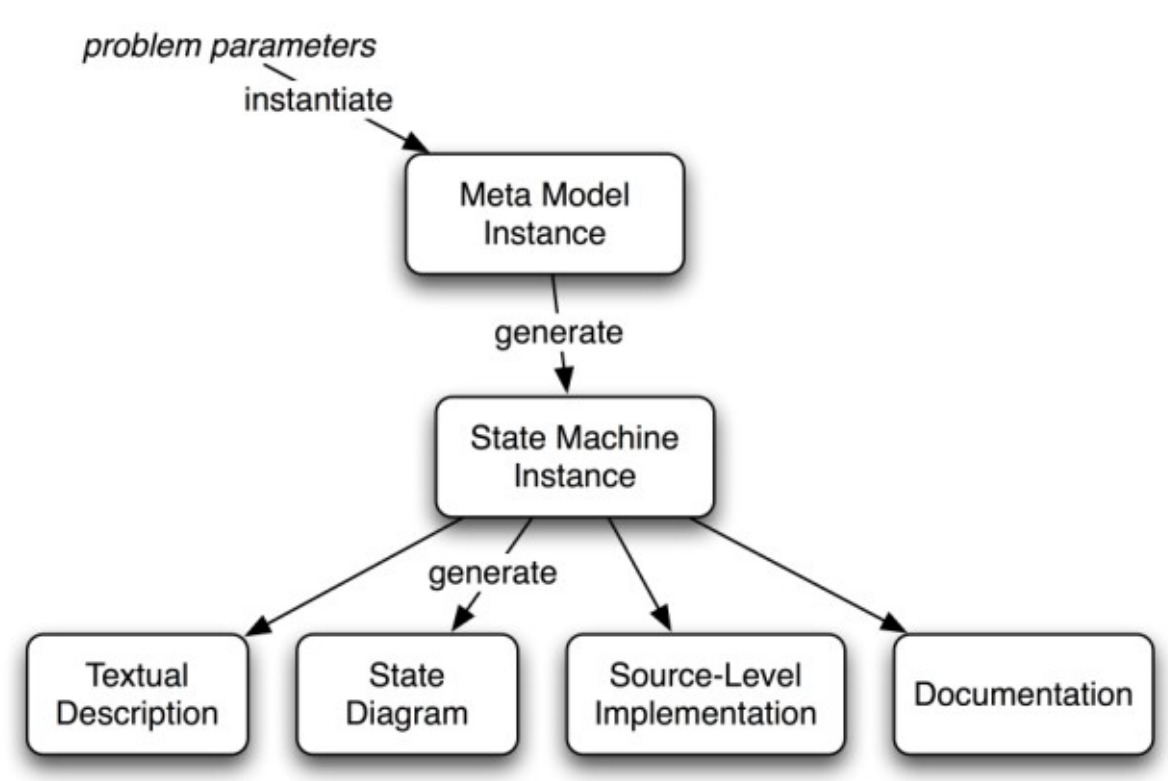

**Figure 1. State machine generation scheme**

The meta-model describes the components of the states, the rules for state update on message receipt, and the actions to be carried out when particular state transitions occur. The meta-model is implemented in Java by a class *MetaModel*. Its constructor takes the replication factor as a parameter, thus each instance of

[1] Details are available at *http://asa.cs.st-andrews.ac.uk/metamodel/*.

*MetaModel* is specialised to that replication factor. The method *generateStateMachine()* performs the generation of the corresponding state machine. This returns an abstract state machine representation in the form of an instance of class *StateMachine*. The state machine contains a collection of states linked by transitions. Both states and transitions may be annotated for documentation purposes. Transitions also refer to associated actions to be performed by the state machine. These classes are outlined in Figure 2.

```
class MetaModel {
        MetaModel(int replication_factor) {…
        StateMachine generateStateMachine();
}
class StateMachine {
        String[] messages;
        State[] states;
        State start_state;
        State finish_state;
}
class State {
        String state_name;
        Transition[] transitions;
        String[] annotations;
}
class Transition {
        State resultant_state;
        String[] actions;
        String[] annotations;
}
```

**Figure 2. Corresponding Java classes**

Figure 3 shows how a particular state machine may be generated and rendered in a textual format.

```
MetaModel meta_model_4 = new MetaModel(4);
StateMachine machine_4 =
        meta_model.generateStateMachine();

println(new TextRenderer().render(machine_4));
```

**Figure 3. Generating a state machine**

## 3.2. Defining the meta-model

In general terms, the meta-model is a model of the structure common to all members of the state machine family. The steps involved in the generation of a particular member of the family—an instance of *StateMachine*—are as follows:

1. generate a data structure containing representations of all possible states
2. for each state, generate the transitions resulting from all possible messages, and record in the data structure
3. prune any unreachable states
4. combine any sets of equivalent states

The final data structure forms the resulting *StateMachine* instance. Of these steps, *1*, *3* and *4* can be performed fairly mechanically, whereas step *2* embodies the core logic of the algorithm.

**3.2.1. Generating possible states.** To generate all possible states, the state space must be defined in terms of the problem parameters—in our case, the replication factor. The BFT commit algorithm involves five messages that may be received by a participating node:

*put*, *vote*, *commit*, *free*, *not free*

In brief, the client sends a *put* message to each of the servers. A *vote* message is sent by a server to the others when it believes that this update should be next in the global ordering. A *commit* message is sent to indicate that enough votes have been received to proceed with the update. The algorithm works by counting the messages sent and received, yielding a state comprising the union of the following variables:

| | |
|---|---|
| ***boolean*** | *put_received, vote_sent, commit_sent* |
| ***int*** | *votes_received, commits_received* |

Two other boolean variables, *could_choose* and *has_chosen*, are used by the algorithm to track the *free* and *not free* messages. The upper bound on both *votes_received* and *commits_received* is one less than the number of participants, which itself is given by the replication factor. Thus in total there are five boolean variables and two integer variables that range from 0 to *r-1* for replication factor *r*. Hence the space of possible states, containing all combinations of values, has the size $2^5r^2$. This gives 512 states for the smallest sensible value of *r=4*. The *generateStateMachine()* operation iterates through all of these combinations, generating a list of *State* objects. A simplified example of the data structure at this stage is shown in Figure 4.

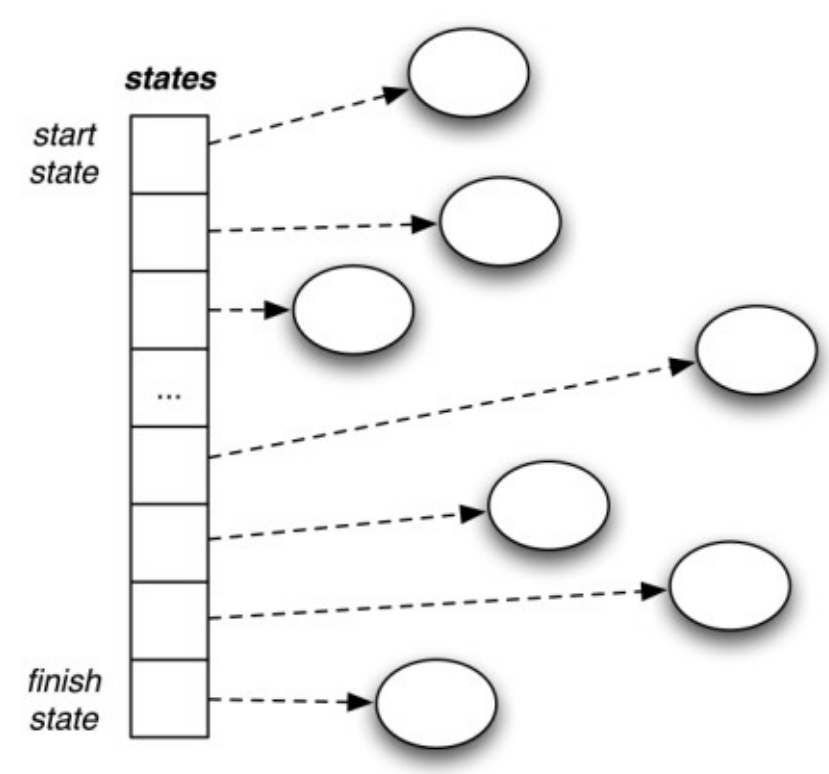

**Figure 4. Data structure after step 1**

**3.2.2. Generating transitions.** The core of the meta-model defines the transitions between states. For any given state, it determines the effects of each of the possible messages, in terms of actions performed and the resulting state. Given that a transition from one state to another represents a change in the variables tracking the messages sent and received, a transition can be categorised as either a simple state transition or a phase transition.

On a simple state transition, the sole effect is to increment one of the received message counts; no action is performed. A phase transition occurs when the receipt of a message causes some threshold to be crossed, triggering an action. For example, in the commit algorithm, when the total number of votes sent and received reaches the number of non-faulty nodes, a *commit* message is sent to all the nodes.

The second step in the generation of a state machine is to iterate over each of the state representations in the data structure generated during the first step. For each state, the meta-model determines which transitions would result from each of the possible messages, if received by the running state machine in that state. Each transition, along with any corresponding actions, is recorded in the state machine data structure.

Figure 5 shows the operation *generateTransitionOnVote()*, defined within the meta-model, determining the transitions from any given state on receipt of a *vote* message[2]. The control decisions that would be taken dynamically in a generic algorithm are here being taken at generation time.

```
generateTransitionOnVote(State s) {
    initialise state variables from s
    increment votes_received
    if total votes >= threshold(r):
        if !vote_sent:
            if could_choose:
                set has_chosen
                record action:
                    send not free message
            record action: send vote message
            set vote_sent
            unset could_choose,
        if commit_sent:
            record action: send commit message
            set commit_sent
    derive new state s1 from state variables
    record transition s->s1 in data structure
}
```

**Figure 5. Meta-model for *vote* message**

Figure 6 shows the data structure after representations of the state transitions have been generated.

**3.2.3. Pruning unreachable states.** Once the complete transition graph has been generated, a reachability analysis is performed. Depending on the application, there may exist states that could never be reached via transitions from the start state. For example, the commit algorithm completes as soon as f+1 commit messages have been received, thus there are no reachable states where the commit count exceeds f. For simplicity, such states are removed from the generated model. With a replication factor of 4, this step reduces the state space from its initial size of 512 to 48.

[2] Similar logic in the meta-model generates documentation describing the states and the rationale for each transition.

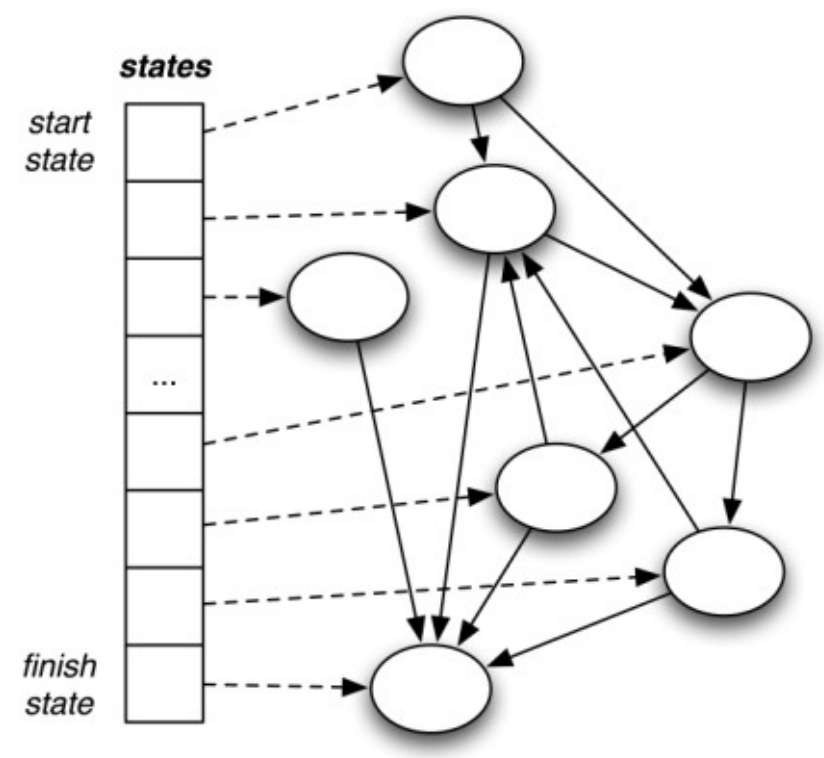

**Figure 6. Data structure after step 2**

Figure 7 illustrates the result of pruning.

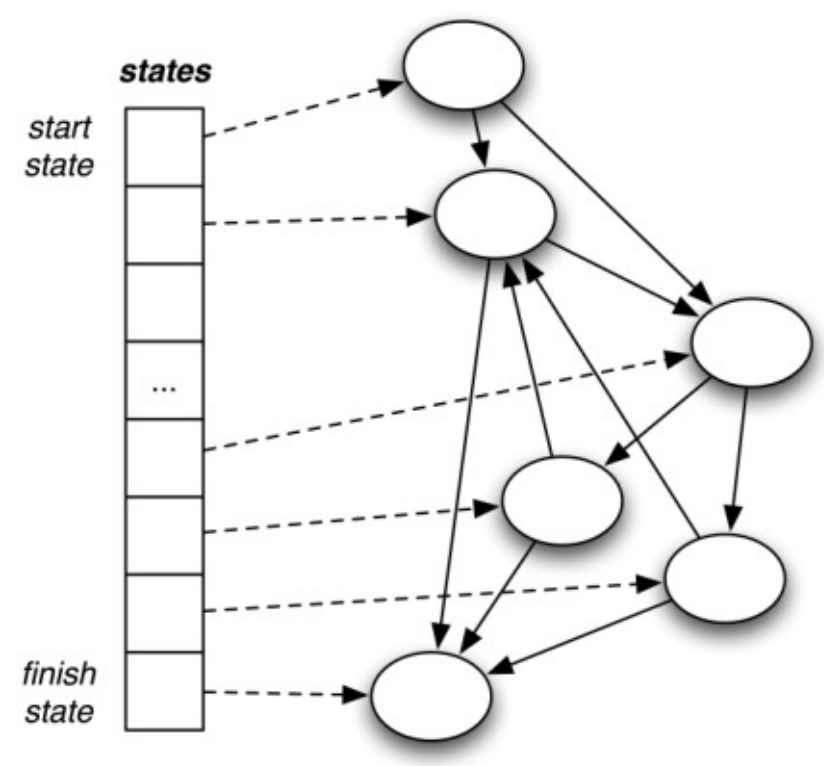

**Figure 7. Data structure after step 3**

**3.2.4. Combining equivalent states.** The generated state machine may be further simplified by identifying and combining sets of states that are equivalent, in the sense that the outgoing transitions from each perform the same actions and lead to the same destination state. Since this step may result in further states becoming unreachable, the previous step and this one are repeated alternately until no further reduction in the state space occurs. With a replication factor of 4, this process eventually results in 33 states. Figure 8 illustrates the result of this step.

## 3.3. State machine artefacts

The abstract representation of a state machine generated by the meta-model can be rendered to yield various concrete artefacts, including:

- a simple textual representation
- a state transition diagram
- source code for an implementation of the corresponding protocol

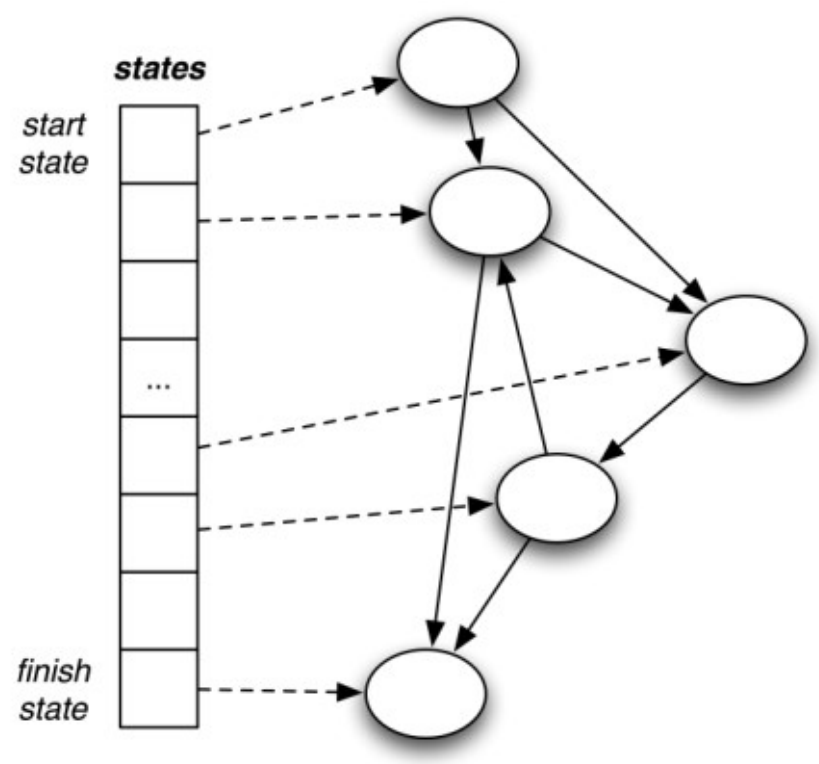

**Figure 8. Data structure after step 4**

Figure 9 shows the textual representation of a particular state and its outgoing transitions. The name of the state encodes the variable values (*put_received*, *votes_sent* etc) in that state. Note that the commentary describing the state in terms of the generic algorithm has been entirely automatically generated.

```
state: T/2/F/0/F/F/F

Have received initial put from client. Have not
voted since another update has already been voted
for. Have received 2 votes and no commits. Have not
sent a commit since neither the vote threshold (3)
nor the external commit threshold (2) has been
reached. May not choose since another ongoing update
has been voted for. Have not chosen this update
since another ongoing update has been chosen. Wait-
ing for 1 further vote (including local vote if any)
before sending commit. Waiting for 2 further exter-
nal commits to finish.

Transitions:

       message: VOTE
           action: send vote message
           action: send commit message
           transition to: T/3/T/0/T/F/F

       message: COMMIT
           transition to: T/2/F/1/F/F/F

       message: FREE
           action: send vote message
           action: send commit message
           action: send not free message
           transition to: T/2/T/0/T/T/T
```

**Figure 9. Example generated state description**

A state machine may be rendered as a state diagram by generating an XML diagram representation that can be imported into a diagramming tool (in this case, Together [4]). An example is available online at the address given previously.

Figure 10 shows a fragment of generated code, dealing with the receipt of a *vote* message. Each state is represented by a generated variable of the form *S-F-0-F-0-F-F-F*. Although the structure embodied in the generated code is equivalent to that shown in Figure 9, its organisation differs in that all possible states are grouped under each message, rather than vice-versa.

```
void receiveVote() {

  switch (getState()) {

    case (S-F-0-F-0-F-F-F) : {
      setState(S-F-1-F-0-F-F-F);
    }

    case (S-F-0-F-0-F-F-T) : {
      setState(S-F-1-F-0-F-F-F);
    }
    ...
    case (S-T-1-T-1-F-T-T) : {
      sendCommit();
      setState(S-T-2-T-1-T-T-T);
    }
    ...
}
```

**Figure 10. Example generated source code**

Commentary on states and transitions, as illustrated in Figure 9, is also included in the generated code.

## 4. Use in practice

We have incorporated the meta-model for the distributed commit algorithm into the ASA infrastructure. Since the replication factor is expected to change only rarely, we executed the meta-model with the default replication factor, generated source code from the resulting state machine, and copied that into the codebase. The benefit of this approach is that the implementation is tightly coupled with its state machine description, giving us confidence in its correctness.

Should we wish in future to support dynamic change to the replication factor, this may be achieved by dynamically generating implementations, compiling them and loading the resulting classes [5]. So long as new replication factors are not presented at high frequency, this approach is quite feasible; Table 1 shows approximate wall-clock times taken to generate state machines of various complexities on an Apple MacBook Pro (3GB, 2.33GHz Intel Core 2 Duo).

**Table 1. Times to generate state machines**

| *f* | *r* | initial states | final states | generation time (s) |
|---|---|---|---|---|
| 1 | 4 | 512 | 33 | 0.10 |
| 2 | 7 | 1568 | 85 | 0.12 |
| 4 | 13 | 5408 | 261 | 0.38 |
| 8 | 25 | 20000 | 901 | 2.2 |
| 15 | 46 | 67712 | 2945 | 19.1 |

Since completing the meta-modelling process for the ASA distributed commit algorithm we have refined the infrastructure to make it applicable to other problems. As steps 1, 3 and 4 described in section 3.2 are largely independent of the details of the algorithm, the implementation of these steps was separated into an abstract superclass, from which problem-specific meta-models can be derived. Rather than containing hard-wired definitions of the state components and messages, these are now represented by a data structure with which the generic meta-model is initialised. Figure 11 shows how the original meta-model is now configured. Each instance of *IntComponent* defines the maximum value of the corresponding state component.

```
StateComponent[] state_components = {
   new IntComponent("votes_received",
        replication_factor - 1),
   new IntComponent("commits_received",
        replication_factor - 1),
   new BooleanComponent("put_received"),
   new BooleanComponent("vote_sent"),
   new BooleanComponent("commit_sent"),
   new BooleanComponent("could_choose"),
   new BooleanComponent("has_chosen")};

String[] messages = {"put", "vote",
   "commit", "free", "not_free"};

initMetaModel(state_components, messages);
```

**Figure 11. Initialising generic meta-model**

## 5. Related work

This work is obviously strongly related to the extensive literature on finite state machines, for example [1, 6]. Traditionally, state machines are used to model computations with fixed numbers of states, whereas our generative approach allows greater flexibility.

Architectural style languages [7, 8] allow families of related systems to be characterised in terms of their shared high level system structure, and specialised to produce particular instances. The work described here is less general since it focuses explicitly on the state machine paradigm; the generic meta-model could be thought of as one particular architectural style.

We have previously used generative techniques to build generic object browsers [9] and to support highly generic strongly typed code [5].

An alternative strategy is to apply formal specification and verification techniques to fault-tolerant algorithms. For example, in [10] a protocol is specified as logical assertions and verified using an interactive proof checker. In [11] an extended actor algebra is used to specify fault-tolerant software architectures. These approaches offer the possibility of formal proofs, whereas here we intend to provide a less formal aid to understanding, at significantly lower cost.

## 6. Conclusions

We have outlined an approach to generating a family of related state machines and corresponding protocol implementations from a unifying meta-model. In the ASA project this has allowed us to produce a state machine style description of our original BFT distributed commit algorithm. This has increased our confidence in the correctness of the algorithm; indeed several errors in the original version were identified during the process. We are currently investigating possibilities for performing more rigorous checking on the state machine formulation.

Although we have applied this approach to a specific BFT distributed algorithm, the approach should be applicable to other critical infrastructure problems involving protocols where the number of states is dependent on a set of parameters.

## 7. Acknowledgments

This work was supported by EPSRC grant GR/S44501/01 and by a Royal Society of Edinburgh / Scottish Executive Support Research Fellowship. Markus Tauber and Rob MacInnis contributed to the development of the distributed commit algorithm.